# Magnetic ground state of FeSe

Qisi Wang,[1,†] Yao Shen,[1,†] Bingying Pan,[1,†] Xiaowen Zhang,[1] K. Ikeuchi,[2] K. Iida,[2] A. D. Christianson,[3,4] H. C. Walker,[5] D. T. Adroja,[5] M. Abdel-Hafiez,[6,7] Xiaojia Chen,[6] D. A. Chareev,[8,10] A. N. Vasiliev,[9,10,11] and Jun Zhao*[1,12]

[1] *State Key Laboratory of Surface Physics and Department of Physics, Fudan University, Shanghai 200433, China*

[2] *Research Center for Neutron Science and Technology, Comprehensive Research Organization for Science and Society, Tokai, Ibaraki 319-1106, Japan*

[3] *Quantum Condensed Matter Division, Oak Ridge National Laboratory, Oak Ridge, Tennessee 37831, USA*

[4] *Department of Physics and Astronomy, University of Tennessee, Knoxville, Tennessee 37996, USA*

[5] *ISIS Facility, Rutherford Appleton Laboratory, STFC, Chilton, Didcot, Oxon OX11 0QX, United Kingdom*

[6] *Center for High Pressure Science and Technology Advanced Research, Shanghai, 201203, China*

[7] *Faculty of Science, Physics Department, Fayoum University, 63514 Fayoum, Egypt*

[8] *Institute of Experimental Mineralogy, Russian Academy of Sciences, 142432 Chernogolovka, Moscow District, Russia*

[9] *Low Temperature Physics and Superconductivity Department, M.V. Lomonosov Moscow State University, 119991 Moscow, Russia*

[10] *Institute of Physics and Technology, Ural Federal University, 620002 Ekaterinburg, Russia*

[11] *National University of Science and Technology "MISiS", Moscow 119049, Russia*

[12] *Collaborative Innovation Center of Advanced Microstructures, Nanjing, 210093, China*

† These authors contribute equally to this work.
* Correspondence and requests for materials should be addressed to J.Z. (zhaoj@fudan.edu.cn).

**Elucidating the nature of the magnetism of a high-temperature superconductor is crucial for establishing its pairing mechanism. The parent compounds of the cuprate and iron-pnictide superconductors exhibit Néel and stripe magnetic order, respectively. However, FeSe, the structurally simplest iron-based superconductor, shows nematic order ($T_s$ = 90 K), but not magnetic order in the parent phase, and its magnetic ground state is intensely debated. Here, we report inelastic neutron-scattering experiments that reveal both stripe and Néel spin fluctuations over a wide energy range at 110 K. On entering the nematic phase, a substantial amount of spectral weight is transferred from the Néel to the stripe spin fluctuations. Moreover, the total fluctuating magnetic moment of FeSe is ~ 60% larger than that in the iron pnictide BaFe$_2$As$_2$. Our results suggest that FeSe is a novel $S$ = 1 nematic quantum-disordered paramagnet interpolating between the Néel and stripe magnetic instabilities.**

Recently, FeSe has attracted considerable interest because of its atypical magnetism[1-6] and fascinating superconducting properties[7-10]. Although the superconducting transition temperature $T_c$ of bulk FeSe ($T_c \approx$ 8 K) is low, it increases drastically under pressure ($T_c \approx$ 37 K; ref. 7), by carrier doping ($T_c \approx$ 40-48 K; refs 8, 9), or in the mono layer limit ($T_c$ ~ 65-109 K, ref. 10). The unique superconducting properties of FeSe are presumably related to its magnetism, which has also been shown to be uncommon. FeSe displays nematic but not stripe magnetic order[1,11] which is unexpected because nematic order has been argued to be the consequence of stripe magnetic order, and both break the $C_4$ lattice symmetry[11]. In iron pnictides, the stripe magnetic order invariably occurs at or immediately below the nematic (tetragonal-to-orthorhombic) ordering temperature[11]. Although previous works have shown that the nematic order could be driven by spin fluctuations without the requirement of magnetic order[2-4,12], the microscopic origin of the absence of the long-range stripe magnetic order in FeSe remains elusive.

Theoretical studies have suggested that the stripe magnetic order in FeSe is absent due to the development of other competing instabilities[2-6]. Several ground states have been proposed, including Néel order[2], staggered dimer/trimers/tetramers magnetic order[3], pair-checkerboard order[5], spin antiferroquadrupolar order[4] and charge current-density wave order[6]. In experimental studies, neutron scattering measurements showed substantial low energy stripe spin fluctuations in single crystal[12] and powder samples[13]. However, because of the limitations of the $q$-space information that can be obtained for powder measurements and the relatively narrow energy range probed previously, the precise nature of the magnetic ground state remains undetermined; this elucidation requires measurements of the momentum structure of the spin fluctuation spectrum from low energy to the zone boundary over the entire Brillouin zone.

In this paper, we used inelastic neutron scattering to map out the spin fluctuation spectra over the entire Brillouin zone in single crystalline FeSe (see Methods). Our data reveal the coexistence of the stripe and Néel spin fluctuations, both of which are coupled with nematicity. In addition, although the spin-fluctuation bandwidth is lower, the total fluctuating magnetic moment ($\langle m^2 \rangle = 5.19\ \mu_B^2/\text{Fe}$) of FeSe is ~ 60% larger than that in the iron pnictide BaFe$_2$As$_2$ (ref. 14). These findings suggest that FeSe is an $S = 1$ nematic quantum-disordered paramagnet interpolating between the Néel and stripe magnetic instabilities[2].

**Results**

**Momentum and energy dependence of the stripe and Néel spin fluctuations**. Figure 1 shows the constant-energy images of spin fluctuations in the (*H*, *K*) plane. Here, in the high temperature tetragonal phase (110 K), the spin response is strongest at **Q** = (1, 0) (marked by dashed ellipse) at 15 meV (Fig. 1b), which is consistent with previous low energy measurements[12,13]. With an increase in energy (Fig. 1c-j), the spin fluctuations show anisotropic dispersion and counter-propagate mainly along the *K* direction. This is analogous to the stripe spin fluctuations detected in other iron based superconductors[11,14,15]. Most notably, in addition to the stripe spin fluctuations, comparatively weaker but clear scattering appears near (1, 1) (Fig. 1b, dashed circle), which implies the presence of spin fluctuations associated with the Néel magnetic instability (see Methods); different from the anisotropic stripe spin fluctuations, the Néel spin fluctuations are nearly isotropic in the transverse and longitudinal directions (Fig. 1b-g). With an increase in the energy up to approximately 150 meV, the Néel and stripe spin fluctuations overlap and cover a broad area centred at (1, 1) (Fig. 1j). Upon cooling to within the nematic phase (*T* = 4 K), the Néel spin-fluctuation signal weakens considerably and is almost undetectable below 35 meV (Fig. 1k, 1l). On the other hand, the momentum structure of the stripe spin fluctuation is essentially unchanged above and below $T_s$ (Fig. 1k-s).

To further elucidate the spin fluctuations in *E*-**Q** space, we projected the spin fluctuations along the *K* direction near (1, 0) and (1, 1) (Fig. 2). Because the incident neutron beam was parallel to the *c* axis, the energy transfer was coupled with *L*. No *L* modulations were observed from the scattering near (1, 0, *L*) and (1, 1, *L*), which indicates a two-dimensional nature of the magnetism. As shown in Fig. 2a, the stripe spin fluctuations stem from (1, 0), split into two branches at ~35 meV and extend up to above ~150 meV at 110 K. The steeply dispersive Néel spin fluctuations are also visible (green arrowheads). As the temperature is lowered to 4 K, the Néel spin fluctuation exhibits a ~ 30 meV gap while the stripe spin fluctuations below 70 meV are clearly enhanced (Fig. 2b).

To quantify the dispersions and intensities of the stripe and Néel spin fluctuations, we made constant-energy cuts at distinct energies (Fig. 3). As Fig. 3d-i show, at $T = 110$ K, the single peak centred at (1, 0) at 15 meV evolves into a pair of peaks along the $K$ direction at $E \geq 35$ meV. By contrast, the peak position of the Néel spin fluctuation (see green arrowheads) shows little change; it only broadens gradually in wavevector with increasing energy. The Néel spin fluctuation is more clearly visible along the transverse direction (Fig. 3m-r) because of the comparatively weaker influence of the stripe spin fluctuations. Here, double peaks formed due to the dispersion are not seen below 68 meV, because the Néel spin fluctuations are commensurate and steeply dispersive (Fig. 3m-r). At higher energies, the Néel and stripe spin fluctuation spectra merge with each other, and their dispersions cannot be determined unambiguously (Fig. 3a-c, 3j-l). The Néel spin fluctuation becomes featureless at low energies (Fig. 3i, 3r) at 4 K, which agrees with the results shown in the constant-energy and $E$-**Q** images (Fig. 1k, 2b). We attempted to fit both types of spectra concurrently using a linear spin-wave theory for the two neighbour ($J_{1(a/b)}$-$J_2$) or three neighbour ($J_{1(a/b)}$-$J_2$-$J_3$) Heisenberg model, where $J_{1(a/b)}$, $J_2$ and $J_3$ are nearest neighbor (in the *a/b* direction), next nearest neighbour and next next nearest neighbour exchange coupling constants, respectively; but this was unsuccessful mainly because this theory cannot account for the observed strong low energy spin excitations at both (1, 0) and (1, 1).

**Momentum integrated local susceptibility above and below $T_s$.** More insight into the nature of the underlying magnetic ground state and its interaction with the nematicity could be acquired by calculating in absolute units the energy dependence of the momentum integrated local susceptibility $\chi''(\omega)$ above and below $T_s$; as Fig. 4a and 4b show, at $T = 110$ K, the Néel spin fluctuation spectral weight is roughly 26% of that of the stripe spin fluctuation below 52 meV, where the two signals are well separated in $q$-space. Upon cooling to $T = 4$ K, the spectral weight loss for the Néel spin fluctuations is approximately recovered by the enhanced stripe spin fluctuations (red shaded areas), and thus the total local susceptibility $\chi''(\omega)$ does not show a marked change across $T_s$ (Fig. 4c). Moreover, the detailed temperature dependence of the stripe and Néel spin fluctuations show that the spectral weight transfer is clearly coupled with the development of the nematic phase (Fig. 4d). At both 4 and 110 K, the total $\chi''(\omega)$ exhibits a high maximum at ~105 meV and extends up to 220 meV (Fig. 2a, 2b, 4c). This bandwidth is considerably lower than that (~ 340 meV) of the stripe ordered BaFe$_2$As$_2$ (ref. 14), which is very likely due to the competition between the Néel and stripe magnetic instabilities. Clearly, this type of competition also prevents the long-range magnetic order in FeSe. By integrating the spectral weight from low energy to the zone boundary, we determined that the total fluctuating moment at 4 K and 110 K are $\langle m^2 \rangle = (g\mu_B)^2 S(S+1) = 5.19 \pm 0.32$ and $5.12 \pm 0.27$ $\mu_B^2$/Fe, respectively, which are larger than those in the superconducting BaFe$_{1.9}$Ni$_{0.1}$As$_2$ ($\langle m^2 \rangle = 3.2$ $\mu_B^2$/Fe) and stripe ordered BaFe$_2$As$_2$ ($\langle m^2 \rangle = 3.17$ $\mu_B^2$/Fe) (ref. 14). Accordingly, this yields an effective spin of $S \approx 0.74$ in FeSe, which likely corresponds to an $S = 1$ ground state in the presence of itinerant electrons.

**Discussion**

The coexistence of the Néel and stripe spin fluctuations is unexpected because FeSe contains only one type of magnetic ions. This differs from the Mn-doped nonsuperconducting iron-pnictide compound BaFe$_{2-x}$Mn$_x$As$_2$, where the Néel magnetic correlation is induced by the local moments of Mn while the stripe magnetic correlation is induced by Fe, given that pure BaMn$_2$As$_2$ is a local-moment Néel type antiferromagnet[16]. Although density functional theory and dynamical mean field theory failed to reproduce the observed band structure of FeSe (refs 17,18), a random phase approximation calculation with an "engineered" tight-binding band structure[19] predicted the spin fluctuations near (1, 0) and (1, $q$). However, in this phenomenological model[19], the (1, $q$) spin fluctuation is incommensurate and gapless below $T_s$, which is inconsistent with our data. The relatively small spin-fluctuation bandwidth and large fluctuating moment together with the low carrier density[17,18] indicate that the magnetic moments in FeSe are more localised than in iron pnictides.

In the more localised case, we can exclude the previously proposed competing staggered dimer/trimers/tetramers magnetic order[3] and pair-checkerboard order[5]. Here, the competition between the Néel and stripe magnetic instabilities could be instead understood within the framework of a frustrated $J_1$-$J_2$ model. In this model, for an $S = 1$ system, the Néel order is stable for $J_2/J_1 \lesssim 0.525$, whereas the stripe order is the ground state for $0.555 \lesssim J_2/J_1$ (refs 20,21). It was predicted that FeSe would be an $S = 1$ nematic quantum paramagnet in the intermediate coupling region ($0.525 \lesssim J_2/J_1 \lesssim 0.555$), which is characterized by gapped stripe and Néel spin fluctuations[2,21]. This agrees with our data that the majority of the spectral weight is concentrated at relatively high energies (~100 meV) even in the presence of itinerant electrons (Fig. 2a, 2b and 4c). Furthermore, in this scenario, the nematic order is viewed as a "vestigial" order that is retained when the static stripe order is suppressed by quantum fluctuations[2]. The finding that the stripe spin fluctuation carries considerably more spectral weight than the Néel spin fluctuation suggests that the system is indeed closer to stripe rather than to Néel order. On this basis, the temperature evolution of the spin fluctuations that we observed can be explained. As the stripe magnetic order breaks the $C_4$ lattice symmetry, while the Néel order preserves it, the orthorhombic/nematic phase transition might partially lift the magnetic frustration and drive the system toward the stripe ordered phase. Thus, the stripe spin fluctuations are enhanced while the Néel spin fluctuations are suppressed and gapped in the nematic phase. These considerations lead to a natural understanding of the paramagnetic nematic phase in FeSe.

We now discuss the evolution of the magnetism, nematicity and superconductivity in FeSe and its derivatives. Since the Néel spin fluctuation is gapped in the nematic phase, it is unlikely responsible for the electron pairing in bulk FeSe at ambient pressure. However, the competition between the Néel and stripe magnetic instabilities across $T_s$ suggests that the magnetic ground state and superconductivity could be highly tunable. Indeed, it has been shown that high pressure not only enhances superconductivity, but also induces static magnetic order in FeSe (refs 22,23). This is surprising as superconductivity always competes with the static magnetic order in iron pnictides. Further neutron diffraction measurements under pressure are required to clarify the nature of this magnetic order. In addition, the proximity of the Néel magnetic instability of FeSe might also have implications for our understanding of the magnetism in the K dosed[9,24], molecule intercalated[25] and mono layer[10,26] FeSe, because in these heavily electron doped compounds, the nematic order and the hole pockets are absent, and the electron pockets at two adjacent zone edges are connected by the Néel wavevector **Q**= (1, 1). Interestingly, this would be in analogy with the cuprate superconductors in terms of the magnetism and Fermi surface topology[27]. To further elucidate the role of the Néel spin fluctuations in iron-based superconductivity, a detailed study of the pressure/electron-doping dependence of the spin correlations in FeSe would be desirable.

To conclude, we have reported the observation of the Néel spin fluctuations over a wide energy range in an iron-based superconductor. We show that the absence of the long-range magnetic order in FeSe is due to the competition between the Néel and stripe magnetic instabilities. This differs from the parent compounds of the cuprate and iron-pnictide high-temperature superconductors, where only one type of magnetic order is observed. Our findings agree with a theoretical prediction that FeSe is a novel $S = 1$ nematic quantum-disordered paramagnet interpolating between the Néel and stripe magnetic instabilities[2], which indicates a connection between the magnetism of the cuprate- and iron-based superconductors. The experimental determination of the nematic magnetic ground state of FeSe will be extremely valuable in identifying the microscopic mechanism of superconductivity in FeSe-based materials[8-10,24-26].

## Methods

**Sample growth and characterizations**. Our FeSe single crystals were grown under a permanent gradient of temperature (~ 400 °C - 330 °C) in the $KCl\text{-}AlCl_3$ flux[28]. The single crystal X-ray and neutron diffraction refinements on our samples indicated a stoichiometric chemical composition to within the error bars, and no interstitial atoms or impurity phases were observed (see Supplementary Note 1, Supplementary Figs 1-3, and Supplementary Table 1). The specific heat, magnetic susceptibility and resistivity measurements performed on randomly selected FeSe single crystals further demonstrate that our sample is a bulk superconductor without detectable impurities. (Supplementary Note 1, Supplementary Figs 4 and 5).

**Neutron scattering experiments**. Our inelastic neutron scattering measurements were carried out on the ARCS time-of-flight chopper spectrometer at the Spallation Neutron Source of Oak Ridge National Laboratory, USA, 4SEASONS chopper spectrometer at the Japan Proton Accelerator Research Complex (J-PARC), and MERLIN chopper spectrometer at the Rutherford Appleton Laboratory, Didcot, UK. The large detector arrays on these instruments allowed us to measure spin excitations over a wide range of energy and momentum. The |**Q**|-dependent background is subtracted for the data below the aluminium phonon cutoff energy of ~ 40 meV (Supplementary Note 2, Supplementary Fig. 7). In order to facilitate comparison with theory and previous measurements, our data were normalised into absolute units by using the elastic incoherent scattering of a standard vanadium sample. The absolute intensity of the resonance mode is consistent with previous low energy data[12] normalized with acoustic phonons (Supplementary Fig. 6). The incident neutron beam was aligned parallel to the $c$ axis. The wavevector **Q** at $(q_x, q_y, q_z)$ is defined as $(H, K, L) = (q_x a/2\pi, q_y b/2\pi, q_z c/2\pi)$ in the reciprocal lattice units in the orthorhombic unit cell. In this unit cell, the magnetic wavevectors associated with the stripe and Néel magnetic order are **Q** = (1, 0) and **Q** = (1, 1), which correspond to the ordering wavevectors of the parent compounds of the iron pnictides and the cuprates, respectively.

**Acknowledgements**

We gratefully acknowledge H. Cao for neutron diffraction experimental support, and Q. Si, R. Yu, A. Kreisel, P. Hirschfeld, B. M. Andersen and R. Valenti for discussions. This work was supported by the Ministry of Science and Technology of China (973 project: 2015CB921302), the National Natural Science Foundation of China (No. 11374059), and the Shanghai Pujiang Scholar Program (No. 13PJ1401100). Research at ORNL's Spallation Neutron Source was sponsored by the Scientific User Facilities Division, Office of Basic Energy Sciences, US Department of Energy. A.N.V. was supported in part by the Ministry of Education and Science of the Russian Federation in the framework of the Increase Competitiveness Program of NUST "MISiS" (No. 2-2014-036). D.A.C. and A.N.V. also acknowledge the support of the Russian Foundation for Basic Research through Grants 13-02-00174, 14-02-92002, 14-02-92693 and Act 211 Government of the Russian Federation (No. 02.A03.21.0006).

**References**

1. McQueen, T. *et al.* Tetragonal-to-orthorhombic structural phase transition at 90 K in the superconductor $Fe_{1.01}Se$. *Phys. Rev. Lett.* **103,** 057002 (2009).

2. Wang, F., Kivelson, S. A. & Lee, D. Nematicity and quantum paramagnetism in FeSe. *Nat. Phys.* **11**, 959-963 (2015).

3. Glasbrenner, J. K. *et al.* Effect of magnetic frustration on nematicity and superconductivity in iron chalcogenides. *Nat. Phys.* **11**, 953-958 (2015).

4. Yu, R. & Si, Q. Antiferroquadrupolar and Ising-nematic orders of a frustrated bilinear-biquadratic Heisenberg model and implications for the magnetism of FeSe. *Phys. Rev. Lett.* **115,** 116401 (2015).

5. Cao, H.-Y., Chen, S., Xiang, H. & Gong, X.-G. Antiferromagnetic ground state with pair-checkerboard order in FeSe. *Phys. Rev. B* **91,** 020504(R) (2015).

6. Chubukov, A. V., Fernandes, R. M. & Schmalian, J. Origin of nematic order in FeSe. *Phys. Rev. B* **91,** 201105(R) (2015).

7. Medvedev, S. *et al.* Electronic and magnetic phase diagram of β-$Fe_{1.01}$Se with superconductivity at 36.7 K under pressure. *Nat. Mater.* **8,** 630–633 (2009).

8. Burrard-Lucas, M. *et al.* Enhancement of the superconducting transition temperature of FeSe by intercalation of a molecular spacer layer. *Nat. Mater.* **12,** 15–19 (2013).

9. Miyata, Y., Nakayama, K., Sugawara, K., Sato, T. & Takahashi, T. High-temperature superconductivity in potassium-coated multilayer FeSe thin films. *Nat. Mater.* **14,** 775–779 (2015).

10. Ge, J.-F. *et al.* Superconductivity above 100 K in single-layer FeSe films on doped $SrTiO_3$. *Nat. Mater.* **14,** 285–289 (2015).

11. Dai, P. C. Antiferromagnetic order and spin dynamics in iron-based superconductors. *Rev. Mod. Phys.* **87,** 855 (2015).

12. Wang, Q. *et al.* Strong interplay between stripe spin fluctuations, nematicity and superconductivity in FeSe. *Nat. Mater.* **15,** 159–163 (2016).

13. Rahn, M. C., Ewings, R. A., Sedlmaier, S. J., Clarke, S. J. & Boothroyd, A. T. Strong (π,0) spin fluctuations in β-FeSe observed by neutron spectroscopy. *Phys. Rev. B* **91,** 180501(R) (2015).

14. Liu, M. *et al.* Nature of magnetic excitations in superconducting $BaFe_{1.9}Ni_{0.1}As_2$. *Nat. Phys.* **8,** 376–381 (2012).

15. Lumsden, M. D. *et al.* Evolution of spin excitations into the superconducting state in $FeTe_{1-x}Se_x$. *Nat. Phys.* **6,** 182–186 (2010).

16. Tucker, G. S. *et al.* Competition between stripe and checkerboard magnetic instabilities in Mn-doped $BaFe_2As_2$. *Phys. Rev. B* **86,** 020503(R) (2012).

17. Watson, M. D. *et al.* Emergence of the nematic electronic state in FeSe. *Phys. Rev. B* **91,** 155106 (2015).

18. Shimojima, T. *et al.* Lifting of xz/yz orbital degeneracy at the structural transition in detwinned

FeSe. *Phys. Rev. B* **90,** 121111(R) (2014).

19. Kreisel, A., Mukherjee, S., Hirschfeld, P. J. & Andersen, B. M. Spin excitations in a model of FeSe with orbital ordering. *Phys. Rev. B* **92,** 224515 (2015).

20. Sushkov, O. P., Oitmaa, J. & Weihong, Z. Quantum phase transitions in the two-dimensional $J_1$-$J_2$ model. *Phys. Rev. B* **63,** 104420 (2001).

21. Jiang, H. C. *et al.* Phase diagram of the frustrated spatially-anisotropic $S=1$ antiferromagnet on a square lattice. *Phys. Rev. B* **79,** 174409 (2009).

22. Bendele, M. *et al.* Coexistence of superconductivity and magnetism in $FeSe_{1-x}$ under pressure. *Phys. Rev. B* **85,** 064517 (2012).

23. Terashima, T. *et al.* Pressure-induced antiferromagnetic transition and phase diagram in FeSe. *J. Phys. Soc. Jpn.* **84,** 063701 (2015).

24. Wen, C. H. P. *et al.* Anomalous correlation effects and unique phase diagram of electron doped FeSe revealed by angle resolved photoemission spectroscopy. *Nat. Commun.* **7,** 10840 (2016).

25. Zhao, L. *et al.* Common Electronic Origin of Superconductivity in (Li,Fe)OHFeSe Bulk Superconductor and Single-Layer $FeSe/SrTiO_3$ Films. *Nat. Commun.* **7,** 10608 (2016).

26. Tan, S. *et al.* Interface-induced superconductivity and strain-dependent spin density waves in $FeSe/SrTiO_3$ thin films. *Nat. Mater.* **12,** 634–640 (2013).

27. Lee, P. A., Nagaosa, N. & Wen, X. G. Doping a Mott insulator: Physics of high-temperature superconductivity. *Rev. Mod. Phys.* **78,** 17 (2006).

28. Chareev, D. *et al.* Single crystal growth and characterization of tetragonal $FeSe_{1-x}$ superconductors. *CrystEngComm* **15,** 1989–1993 (2013).

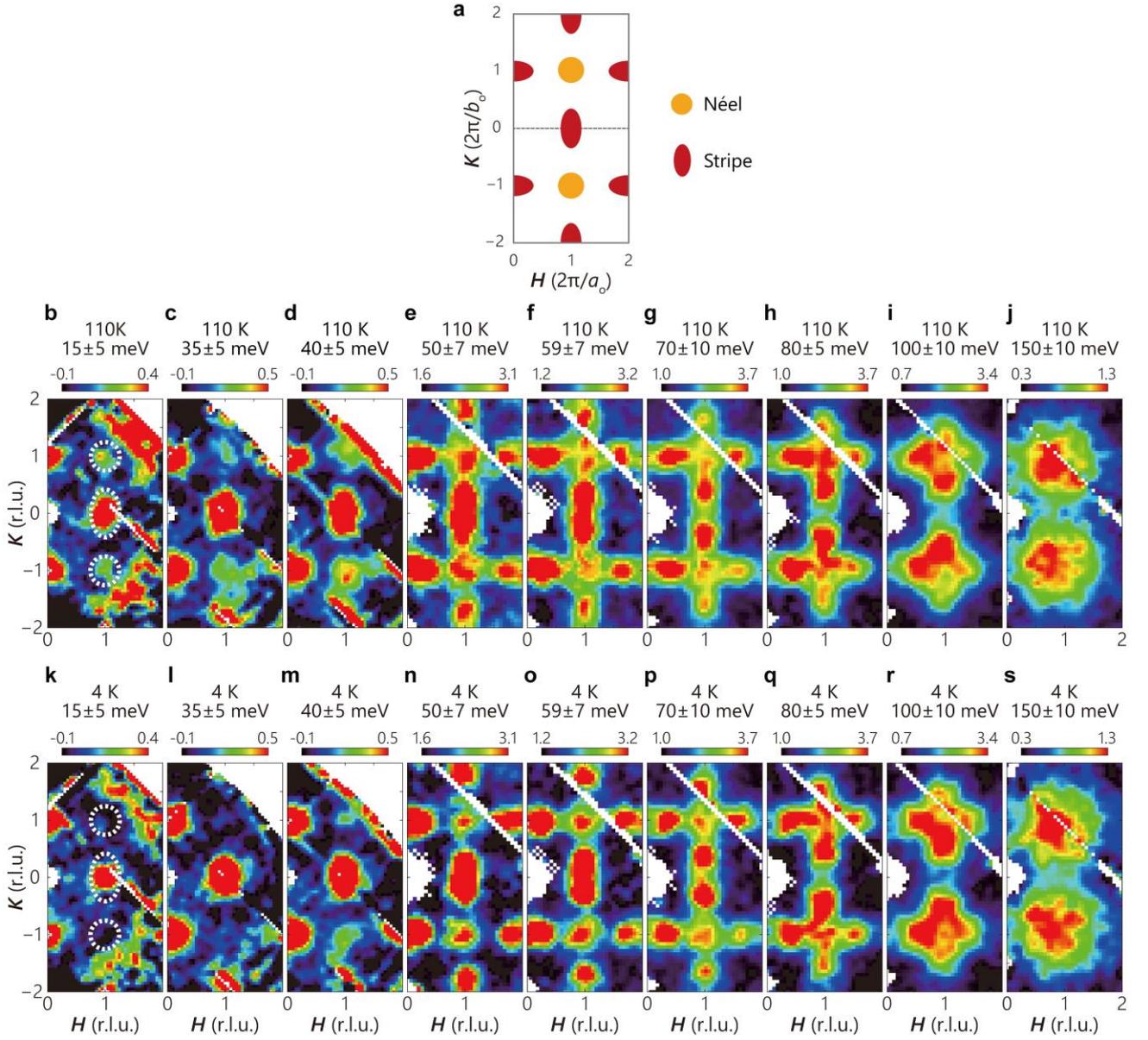

**Figure 1 | Momentum dependence of the spin fluctuations in FeSe at 4 K and 110 K.** (**a**) Schematic representation of the stripe and Néel spin fluctuations in the (*H*, *K*) plane. (**b-j**) Constant-energy images acquired at 110 K at indicated energies. (**k-s**) Constant-energy images obtained at 4 K at the same intensity scale as those acquired at 110 K. The measurements in (**b-d**, **k-m**) and (**e-j**, **n-s**) were carried out on ARCS with the incident neutron energy of 79 and 294 meV, respectively. The sample has two equally populated orthogonal twin domains in the *ab* plane at 4 K and the intensities near (1, 0) and (0, 1) are roughly the same. Symmetry equivalent data were pooled to enhance statistical accuracy. The |**Q**|-dependent background is subtracted for the data (**b-d**, **k-m**) below the aluminium phonon cutoff energy of ~ 40 meV. Above 40 meV, raw data are presented (**e-j**, **n-s**). The colour bars indicate intensity in unit of mbar sr$^{-1}$ meV$^{-1}$ f.u.$^{-1}$. The dashed ellipses and circles mark the stripe and Néel wavevectors, respectively.

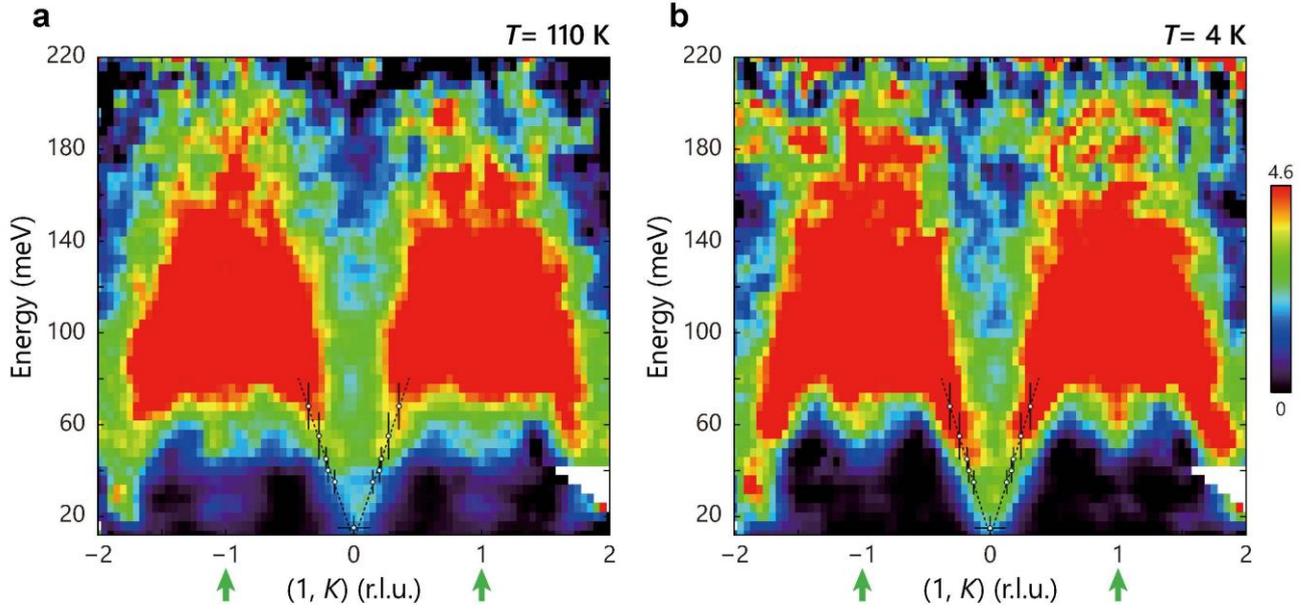

**Figure 2 | Dispersions of the stripe and Néel spin fluctuations in FeSe at 4 K and 110 K.** Background-subtracted *E-K* slice of the spin fluctuations at various incident energies: (**a**) $T$ = 110 K; (**b**) $T$ = 4 K. The data at $E \geq 40$ and $E \leq 40$ meV were collected on ARCS by using incident energy of 294 and 79 meV, respectively. The isotropic $Fe^{2+}$ magnetic form factor is corrected for both sets of data. The spectral weight transfer from the Néel (1, 1) to stripe (1, 0) wavevector below ~70 meV upon cooling to 4 K can be clearly seen. The open circles are dispersions obtained from the constant energy cuts below 68 meV in Fig. 3. The colour bar indicates intensity in unit of mbar sr$^{-1}$ meV$^{-1}$ f.u.$^{-1}$. The vertical bars indicate the energy integration range. The horizontal bars at 15 meV indicate the full width at half maximum of the Gaussian fittings in Fig. 3i. The horizontal bars at other energies are the errors derived by least-square fittings. The dashed lines are a guide to the eye.

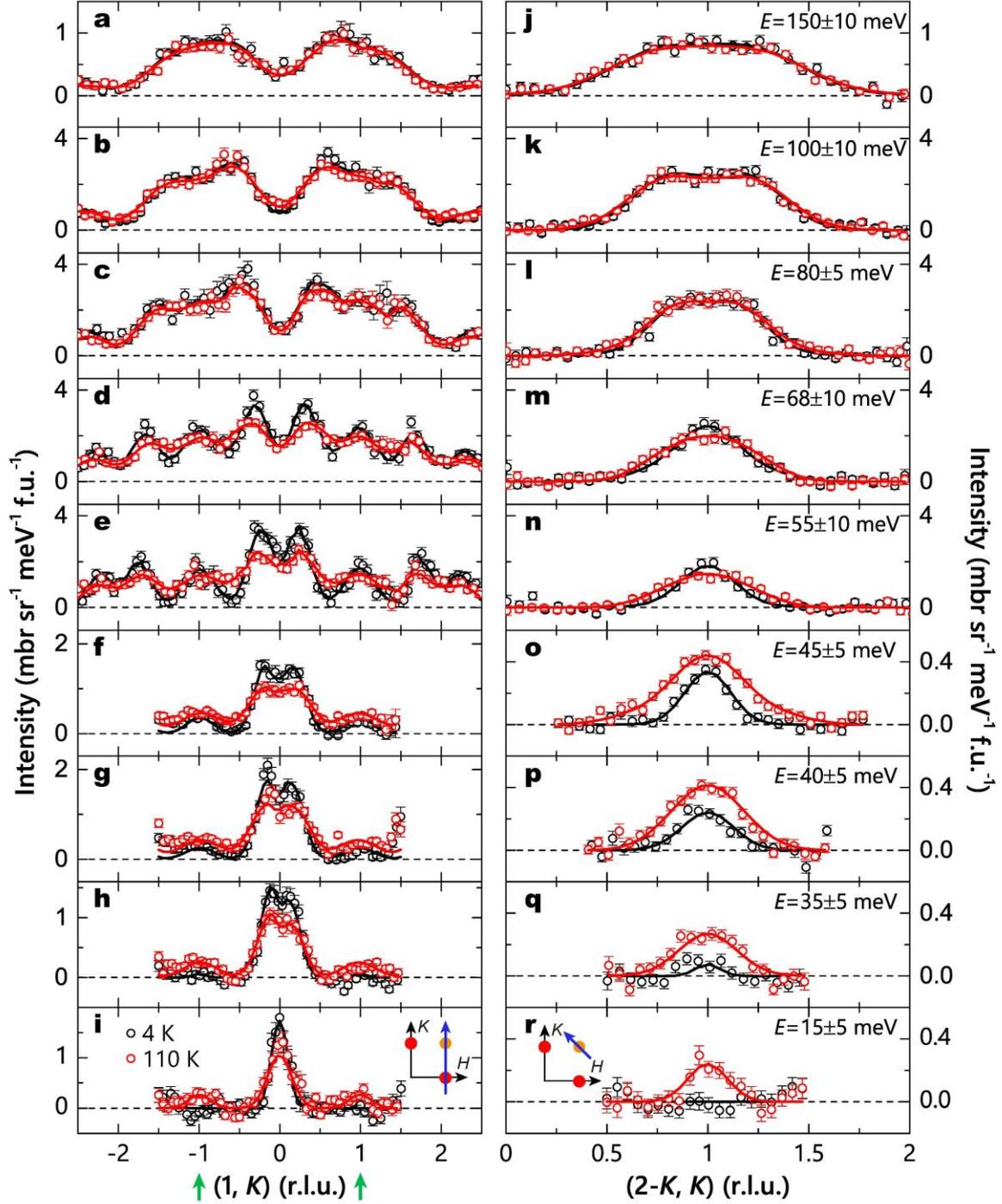

**Figure 3 | Constant-energy cuts of the stripe and Néel spin fluctuations in FeSe at 4 K and 110 K.** (**a-i**) Constant-energy cuts through the stripe and Néel magnetic wavevectors along the $K$ direction at 4 and 110 K. (**j-r**) Constant-energy cuts through the Néel magnetic wavevector $\mathbf{Q} = (1, 1)$ along the transverse direction. The scan directions are marked by the arrows in the insets. The peak positions (dispersions) are determined by fitting with Gaussian profiles convoluted with the instrumental resolution, with the $Fe^{2+}$ magnetic form factor corrected. The fitted dispersions are shown in Fig. 2. The error bars indicate one standard deviation.

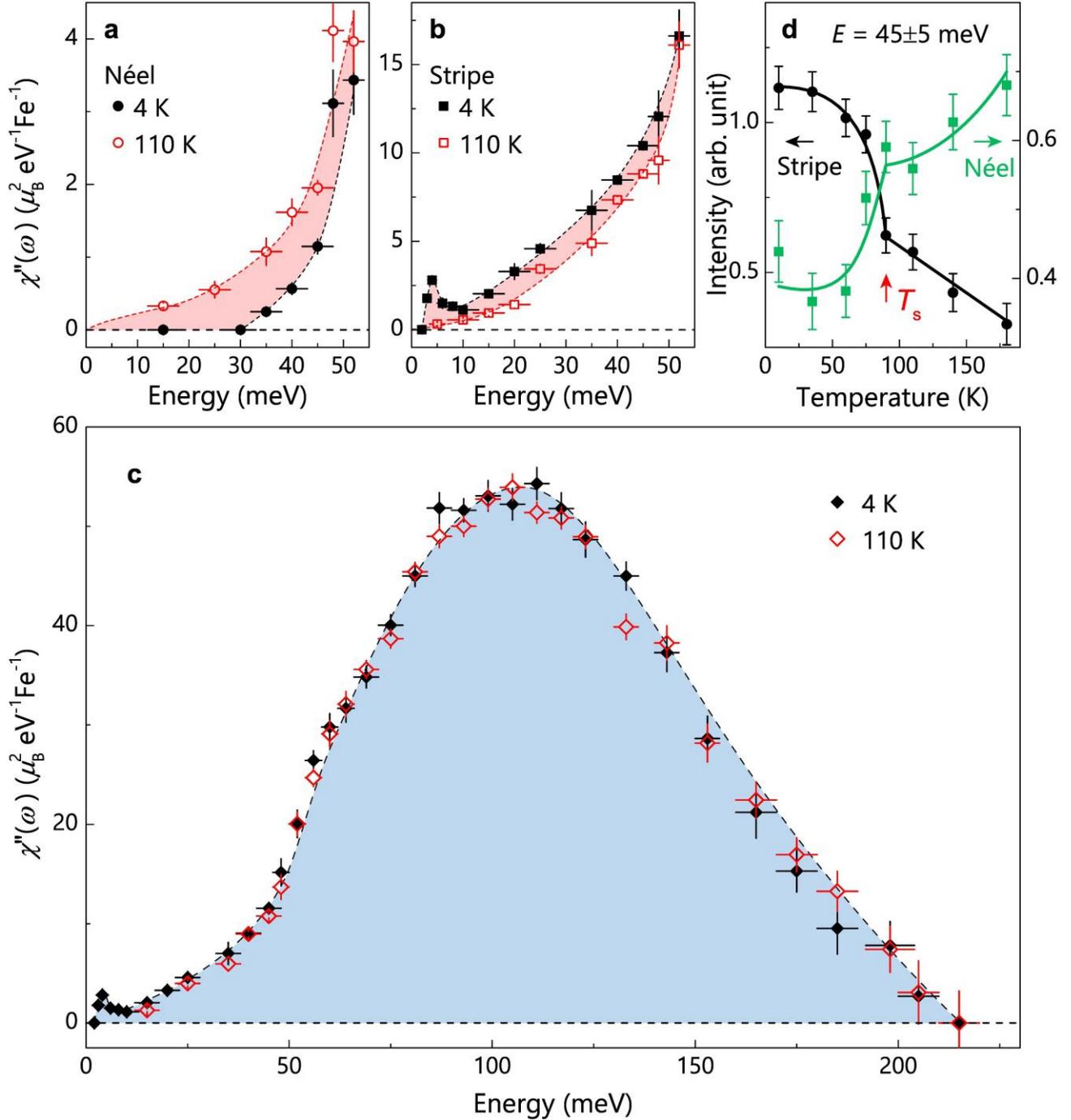

**Figure 4 | Energy dependence of the local susceptibility χ″(ω) for FeSe at 4 K and 110 K.** (**a-c**) Energy dependence of χ″(ω), at 4 and 110 K, calculated for (**a**) the Néel spin fluctuations, (**b**) the stripe spin fluctuations, and (**c**) the sum of the stripe and Néel spin fluctuations. A resonance mode is clearly observed at ~ 4 meV and 4 K, whose intensity is consistent with previous low energy measurements normalized with acoustic phonons[12]. (**d**) Temperature dependence of the intensities of the stripe and Néel spin fluctuations. The data in (**a-c**) were collected on ARCS (incident energy $E_i$ = 294, 79, 40 meV) and 4SEASONS ($E_i$ = 21, 13.6 meV). The data in (**d**) were collected on MERLIN ($E_i$ = 123.4 meV). The horizontal and vertical bars indicate the energy integration range and the statistical errors of one standard deviation, respectively. The solid and dashed lines are a guide to the eye.

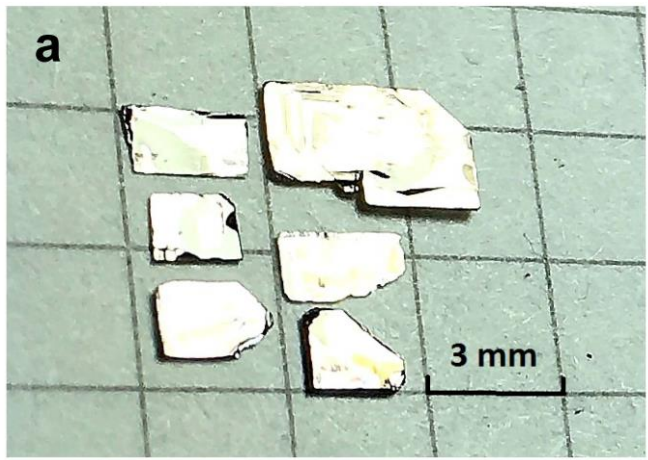 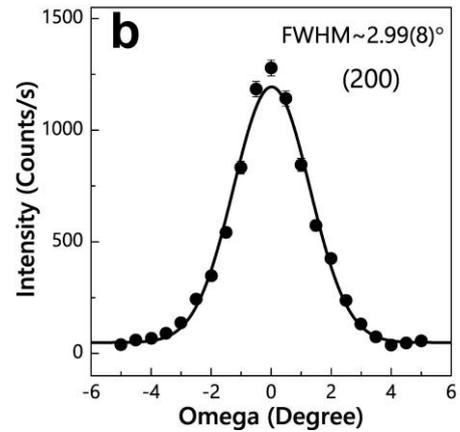

**Supplementary Figure 1: Photograph of our FeSe single crystals and the rocking curve of the co-aligned crystals.** (**a**) Photograph of representative FeSe single crystals (~2-20 mg each) used for inelastic neutron scattering (INS) measurements. The single crystals are plate-like with natural cleaved edges along the tetragonal a/b direction. The well-shaped habitus with metallic luster indicates the high quality of the crystal. (**b**) Rocking curve of the (200) Bragg peak of the co-aligned crystals used for INS measurements. The full width at half maximum (FWHM) of the rocking curve is 2.99(8)°. The error bars indicate one standard deviation.

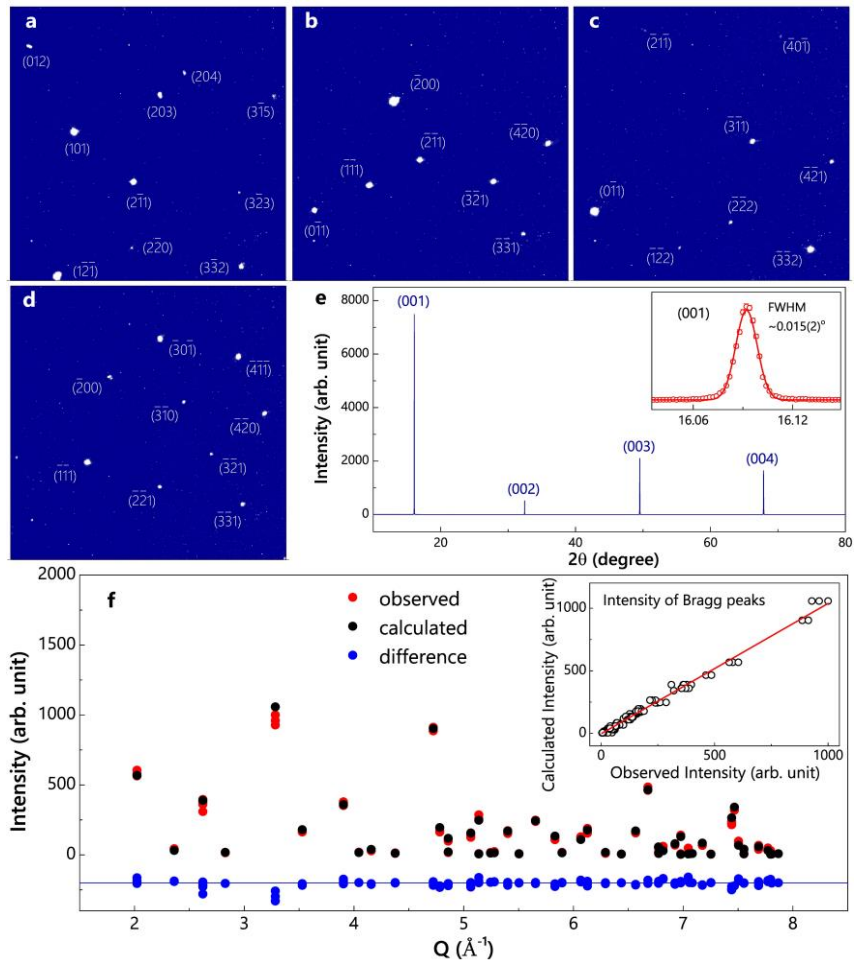

**Supplementary Figure 2: Single crystal X-ray diffraction data and the Rietveld refinement results in FeSe.** (**a**-**d**) Representative diffraction pattern shown on the area detectors at 298 K. The sharp bright spots are indexed as Bragg reflections of the tetragonal phase of FeSe. The measurements were performed on a single crystal from the same batch as those used for INS experiments. (**e**) The scans perpendicular to the cleaved surface (*ab* plane) show only (00L) Bragg peaks of the tetragonal FeSe. The FWHM of the (001) Bragg peak is ~ 0.015(2)°, indicating a good quality of the sample. (**f**) X-ray single crystal diffraction refinement results. The refined parameters are summarized in Table I of the supplementary information of ref. 2. Data are presented in the tetragonal notation. Our Rietveld refinements are based on 178 X-ray Bragg peaks using the FULLPROF program.

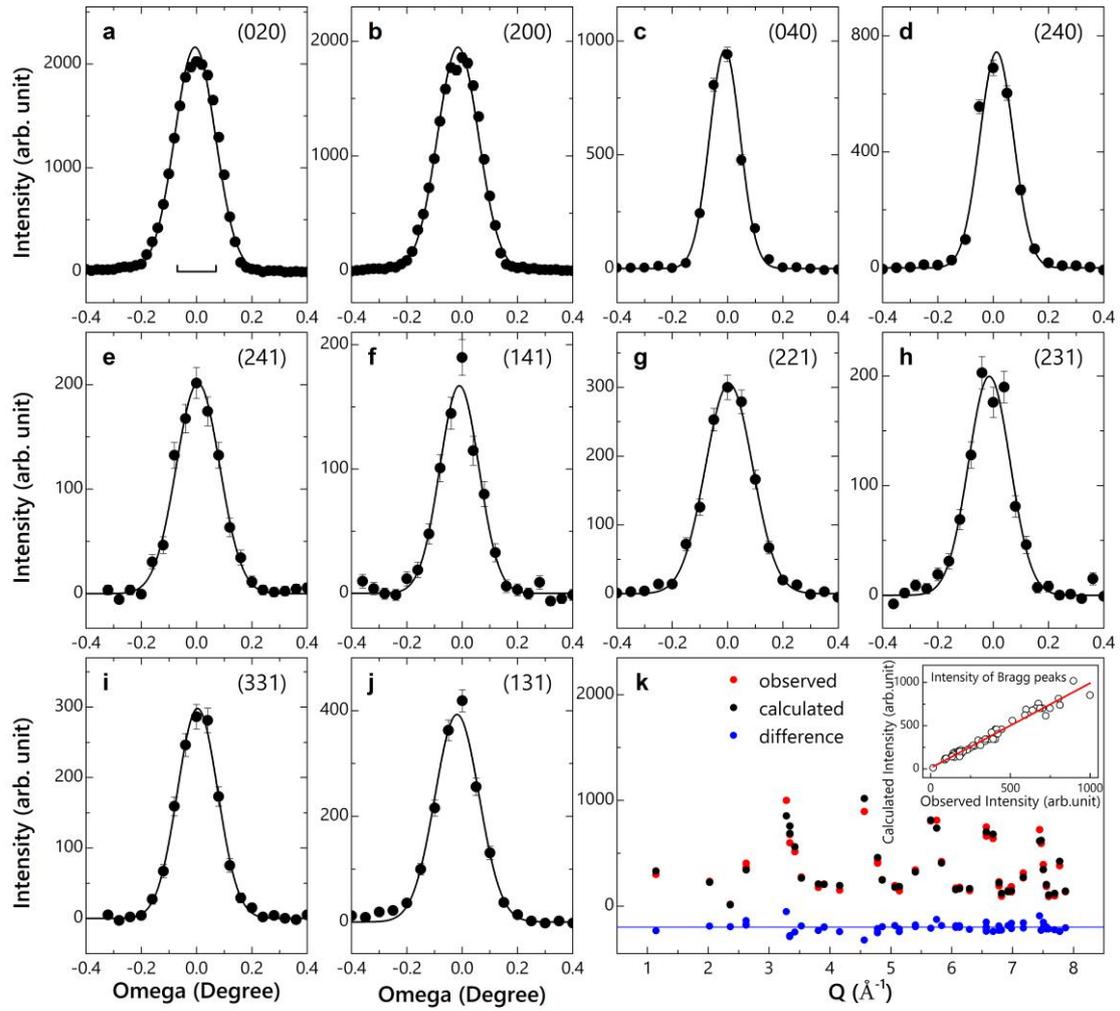

**Supplementary Figure 3: Single crystal neutron diffraction data and the Rietveld refinement results in FeSe.** (**a**-**j**) Representative neutron diffraction pattern measured at 296 K. The FWHM of the rocking curve of (020) peak is 0.173(3)°, which is essentially resolution limited. The horizontal bar indicates the instrument resolution. The measurements were performed on the HB3A four-circle single-crystal diffractometer at the High-Flux Isotope Reactor at the Oak Ridge National Laboratory. (**k**) Single crystal neutron diffraction refinement results. The refined parameters are summarized in Supplementary Table 1. Data are presented in the tetragonal notation. The Rietveld refinements are based on 71 neutron Bragg peaks using the FULLPROF program. The error bars indicate one standard deviation.

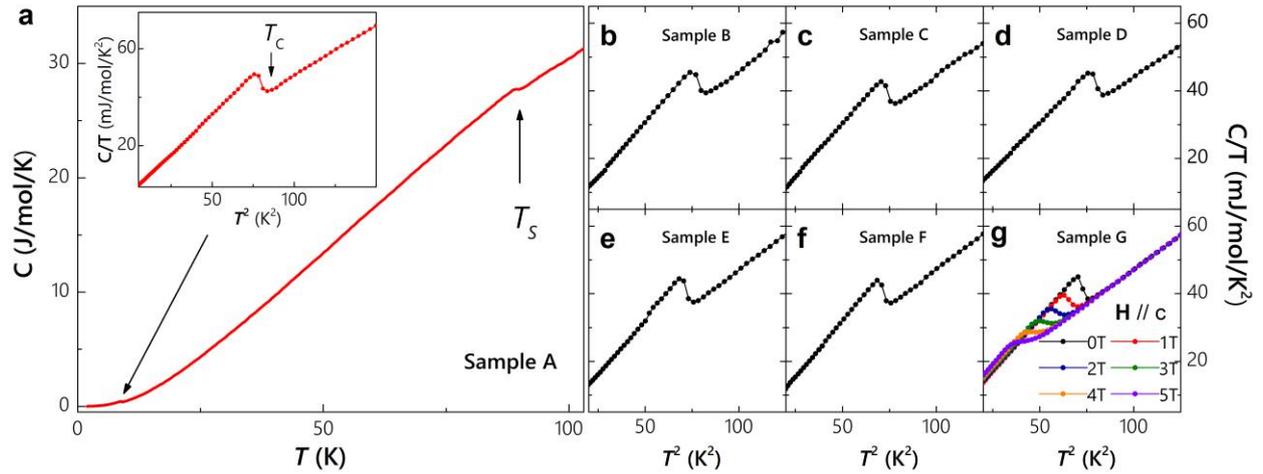

**Supplementary Figure 4: Specific heat of FeSe single crystals.** Samples A-G are randomly selected from the same batch as those used for the INS experiments. (**a**) Temperature dependence of specific heat of sample A. The superconducting and structural transitions are clearly observed. The inset shows the specific heat of sample A near $T_c$. (**b-f**) Low temperature specific heat of samples B-F. Sharp superconducting specific heat anomaly is observed in all samples. (**g**) Magnetic field dependence of the specific heat of sample G. The superconducting specific heat anomaly is progressively suppressed under magnetic field along the *c*-axis.

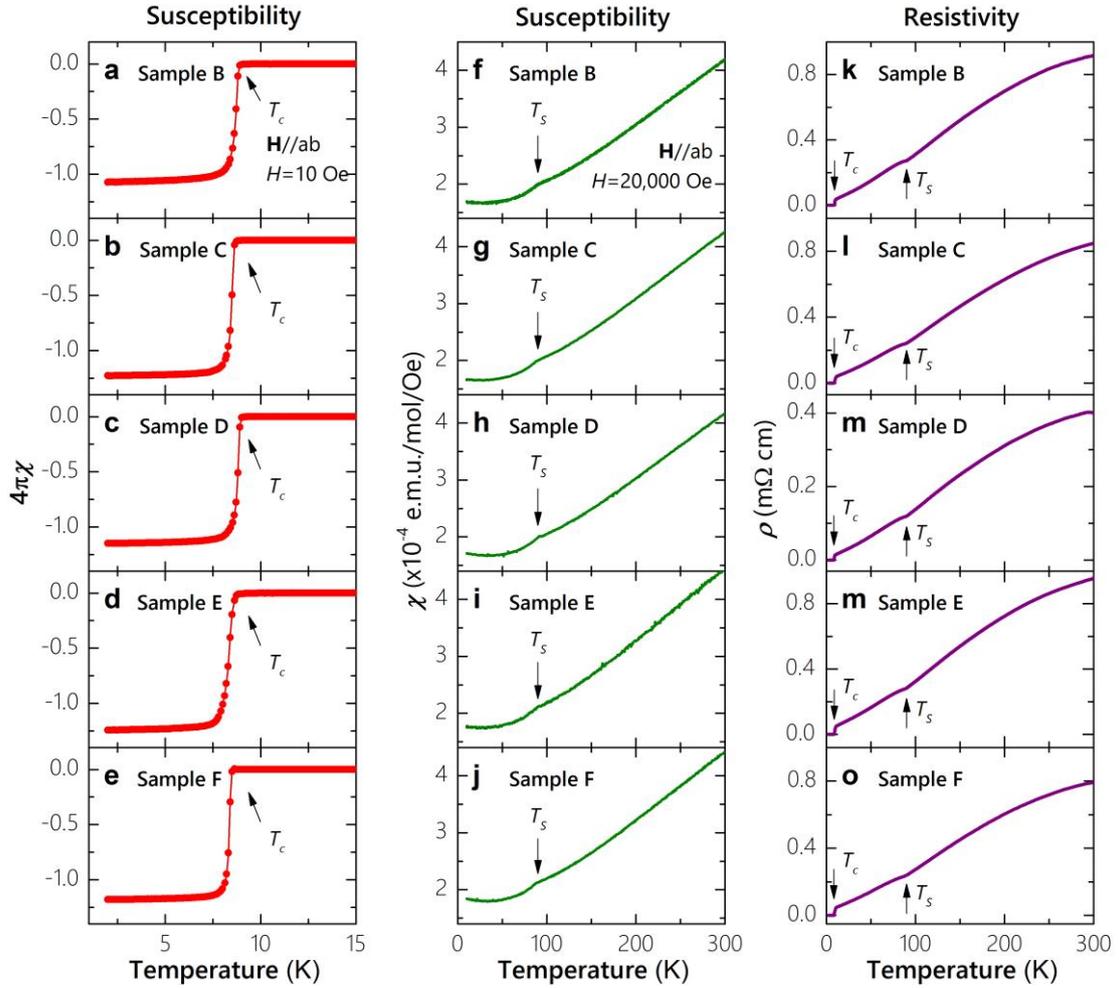

**Supplementary Figure 5: DC magnetic susceptibility and resistivity of FeSe single crystals.** Data were measured on the same samples used for the specific heat measurements. (**a-e**) Zero-field-cooled (ZFC) magnetic susceptibility measured in a magnetic field of $H$=10 Oe. A sharp superconducting transition is observed at ~8.7 K. The superconducting volume fraction is close to ~100% for all samples. The diamagnetic screening is slightly larger than -1 because of the demagnetization effect. (**f-j**) Magnetic susceptibility under a magnetic field of $H$=20000 Oe. The kink at 90 K corresponds to the structural (nematic) phase transition. (**k-o**) Temperature dependence of resistivity. The superconducting and structural transitions are clearly observed.

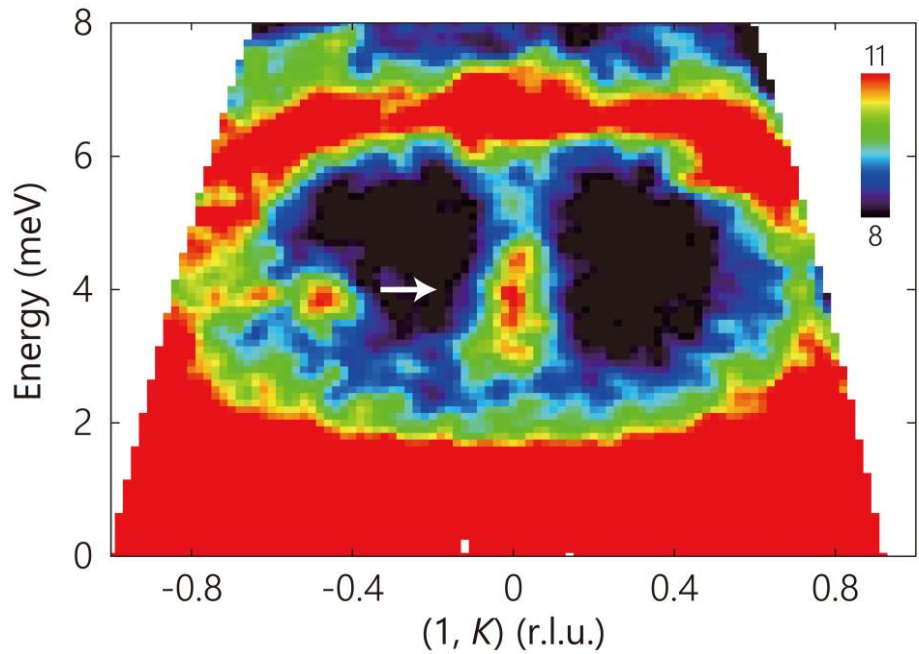

**Supplementary Figure 6: Magnetic resonance mode.** The $E$ - $K$ slice of the spin fluctuations was measured with $E_i$ = 13.6 meV at 4 K on 4SEASONS (ref. 3). A magnetic resonance mode emerges at the stripe AFM wavevector $\mathbf{Q}$ = (1, 0) and $E \approx 4$ meV in the superconducting state (white arrow), which is consistent with our previous data measured on a triple axis spectrometer[2]. By contrast, the Néel spin fluctuation is gapped (up to ~ 30 meV) without a resonance mode at 4 K. The colour bar indicates intensity in arbitrary unit.

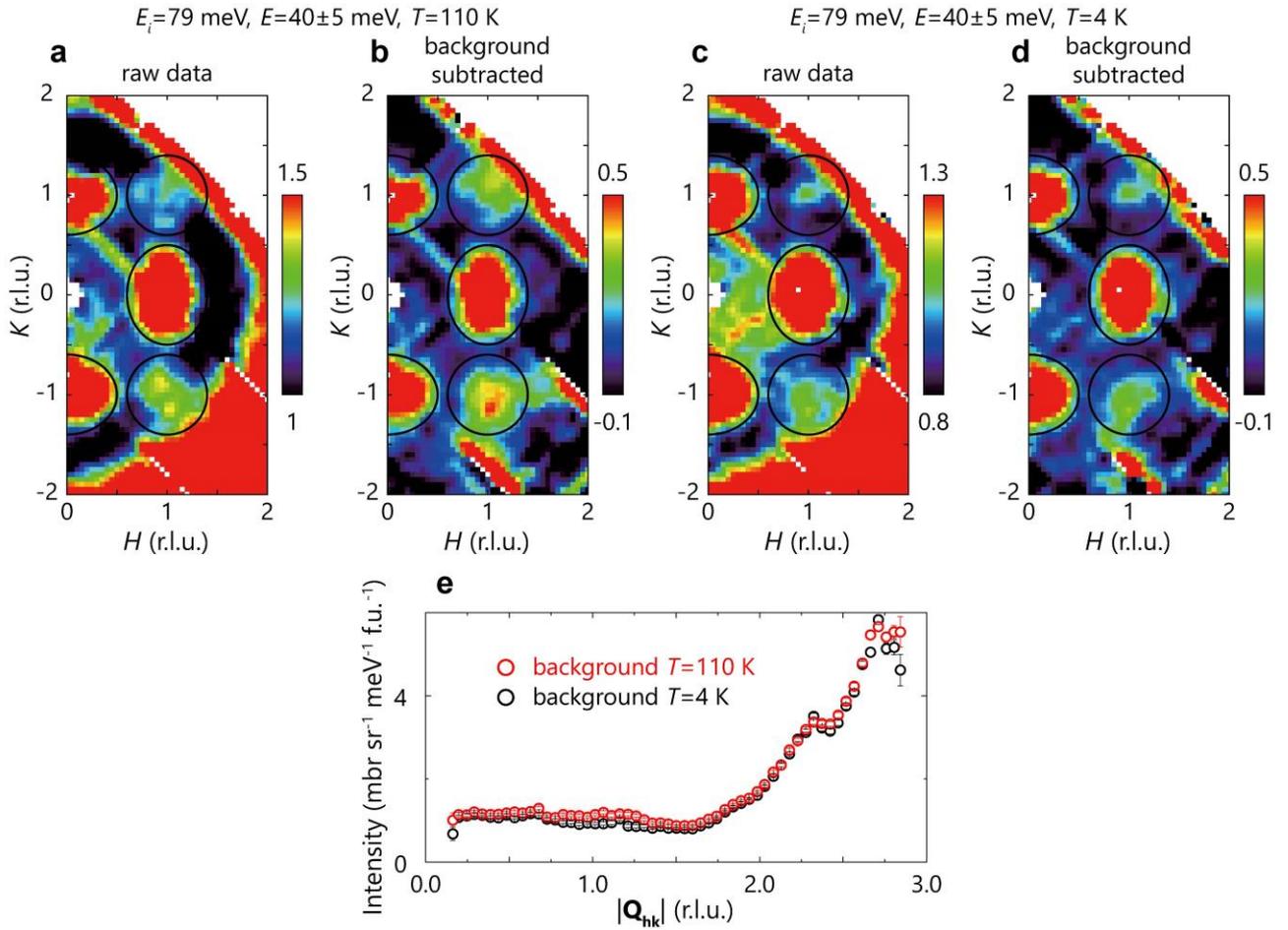

**Supplementary Figure 7: Raw and background-subtracted constant energy images.** (**a, b**) Raw (**a**) and background subtracted (**b**) constant energy images at 40 meV at 110 K. (**c, d**) Raw (**c**) and background subtracted (**d**) constant energy images at 40 meV at 4 K. The black ellipses indicate the magnetic signals. The ring-like phonon background from the aluminium sample holder/environment can be clearly seen in the raw constant energy images (red area at high |**Q**|). The red streaks are due to gaps between neutron detectors. (**e**) The |**Q**|-dependent background estimated from the scattering away from the magnetic signals. The colour bars indicates intensity in unit of mbar sr$^{-1}$ meV$^{-1}$ f.u.$^{-1}$. The error bars indicate one standard deviation.

**Supplementary Table 1.** Refined structure parameters and chemical composition of FeSe against neutron diffraction data at 296 K. Space group: $P4/nmm$ (No. 129). Atomic positions: Fe: 2a (0, 0, 0); Se: 2c (0, 0.5, z).

|  | Refined composition | FeSe$_{0.99(3)}$ |
|---|---|---|
|  | $a$ (Å) | 3.7613(14) |
|  | $c$ (Å) | 5.506(5) |
| Fe atom | $B_{iso}$ (Å$^2$) | 0.64(7) |
| Se atom | $z$ | 0.2660(16) |
|  | $B_{iso}$ (Å$^2$) | 0.95(2) |
|  | $R_1$ | 0.0492 |
|  | $wRF^2$ | 0.118 |
|  | $\chi^2$ | 1.14 |

## Supplementary Note 1: Sample characterizations

To determine the chemical composition and structural parameters of our FeSe single crystals, we have carried out single crystal X-ray and neutron diffraction measurements on single crystals from the same batch as those used for INS experiments. Supplementary Figure 2 shows the single crystal X-ray diffraction data and the Rietveld refinement results. The Rietveld refinements were performed using the FULLPROF program[1]. The refined results are summarized in Table I of the supplementary information of ref. 2. The Bragg reflections are clear and sharp (Supplementary Figs 2a-e), indicating a good crystallization quality, and no scattering from impurity phases or interstitial atoms were observed. The refined chemical composition is FeSe$_{0.990(10)}$, (i.e. stoichiometric within the error bars).

Supplementary Figure 3 presents the representative neutron diffraction patterns and the refinement results. The mosaic of the single crystal determined by the neutron rocking scan of the (020) peak is 0.173(3)°, which is essentially resolution limited (Supplementary Fig. 3a). The neutron diffraction refined chemical composition is FeSe$_{0.99(3)}$ (Supplementary Fig. 3k and Supplementary Table 1), which is consistent with the X-ray diffraction results.

We have performed specific heat, magnetization and resistivity measurements on randomly selected single crystals from the same batch as those used for INS experiments (Supplementary Figs 4 and 5). As Supplementary Figure 4 shows, the specific heat of our crystal exhibits a sharp jump near $T_c$. The magnetization and resistivity measurements give a very sharp superconducting transition at $T_c$ ~ 8.7 K with a transition width of ~ 0.3 K, and the superconducting volume fraction is close to ~100% for all samples (Supplementary Figs 5a-e). We also note that the temperature dependence of the magnetic susceptibility and specific heat on our sample displays no indication of magnetic impurities.

## Supplementary Note 2: Background subtraction

We used an aluminium sample holder/environment for our inelastic neutron scattering measurements because aluminium has a high transparency for neutrons. The phonon background from the polycrystalline aluminium sample holder/environment below its phonon cut-off energy of ~ 40 meV only depends on the amplitude of **Q** (similar to a powder ring), which could be easily subtracted. Supplementary Figure 7a and 7c show the raw data of constant energy images at $E = 40\pm5$ meV at 110 K and 4 K, respectively. A |**Q**|-dependent background was subtracted and the resultant data was shown in Supplementary Figure 7b and 7d in which the stripe and Néel magnetic signals can be seen more clearly. The |**Q**|-dependent background was estimated by the scattering away from the magnetic signals (black ellipses). For the data at $E > 40$ meV, no ring-like phonon background was observed and the raw constant energy images were presented in Fig. 1e-j, 1n-s.

## Supplementary References

[1] Rodrguez-Carvajal, J. Recent advances in magnetic structure determination by neutron powder diffraction. *Physica (Amsterdan)*, **192B,** 55 (1993).

[2] Wang, Q. *et al.* Strong interplay between stripe spin fluctuations, nematicity and superconductivity in FeSe. *Nat. Mater.* **15,** 159-163 (2016).

[3] Kajimoto, R. *et al.* The Fermi Chopper Spectrometer 4SEASONS at J-PARC. *J. Phys. Soc. Japan* **80,** SB025 (2011).